\documentclass[a4paper,10pt,twoside]{article}
\pdfoutput=1
\usepackage{mathpazo}
\usepackage[zerostyle=c]{newtxtt}
\usepackage[T1]{fontenc}
\usepackage{hyphenat}
\usepackage{amsmath}
\usepackage{bm}
\usepackage{booktabs}
\usepackage[hidelinks]{hyperref}
\usepackage{graphicx}
\usepackage[top=3cm,bottom=2.5cm,left=2.5cm,right=2.5cm]{geometry}
\usepackage[labelfont={bf},labelsep=period,justification=justified,hang]{caption}
\usepackage[caption=false,font=footnotesize]{subfig}
\usepackage{xcolor}
\usepackage{ifthen}
\usepackage{fancyhdr}
\usepackage{url}
\usepackage{eso-pic}
\usepackage{etoolbox}
\usepackage{siunitx}
\usepackage{threeparttable}

\graphicspath{{./gfx/}}
\newcommand{\Title}[1]{\def\Title{#1}}
\newcommand{\TitleRunning}[1]{\def\TitleRunning{#1}}
\newcommand{\Author}[1]{\def\Author{#1}}
\newcommand{\AuthorRunning}[1]{\def\AuthorRunning{#1}}
\newcommand{\Address}[1]{\def\Address{#1}}
\newcommand{\Abstract}[1]{\def\Abstract{#1}}

\renewcommand{\maketitle}{%
    \thispagestyle{empty}%
    \begin{center}
    {\Large\bfseries\Title}
    \\[1em]
    {\large\Author}
    \\[0.5em]
    {\Address}
    \\[1em]
    \parbox{0.8\linewidth}{\paragraph*{Abstract}\Abstract}
    \end{center}
    \vspace{1em}
}

\newcommand{\D}{\displaystyle}
\newcommand{\Vector}[1]{\bm{#1}}  % for vectors
\newcommand{\Matrix}[1]{\bm{#1}}  % for matrices
\newcommand{\Transpose}{\mathrm{T}}  % alternatives for transpose symbol: \mathrm{T}, \intercal, \top
\DeclareMathOperator{\diag}{diag}
\newcommand{\refEq}[1]{(\ref{#1})}               % equation
\newcommand{\refEqBegin}[1]{Equation (\ref{#1})} % eq. at beginning of sentence
\newcommand{\refFig}[1]{Fig.~\ref{#1}}         % figure
\newcommand{\refFigBegin}[1]{Figure~\ref{#1}}    % fig. at beginning of sentence
\newcommand{\refTable}[1]{table~\ref{#1}}        % table
\newcommand{\refTableBegin}[1]{Table~\ref{#1}}   % table at beginning of sentence
\renewcommand{\S}{\mathrm{S}}
\renewcommand{\L}{\mathrm{L}}
\newcommand{\SL}{\mathrm{S/L}}

\Title{A Building-Block Approach to State-Space Modeling of \protect\\ DC-DC Converter Systems}
\TitleRunning{A Building-Block Approach to State-Space Modeling of DC-DC Converter Systems}

\AuthorRunning{Gernot Herbst}%
\Author{Gernot Herbst%
    \ \href{https://orcid.org/0000-0002-4638-5378}%
    {\raisebox{-0.3pt}{\includegraphics[height=9pt]{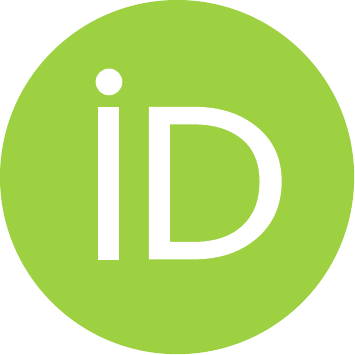}}}%
}
\Address{%
    Siemens AG, Chemnitz, Germany\\%
    \url{gernot.herbst@siemens.com}
}%

\Abstract{%
Small-signal models of DC-DC converters are often based on a state-space averaging approach, from which both control-oriented and other frequency-domain characteristics, such as input or output impedance, can be derived.
Updating these models when extending the converter by filters or non-trivial loads, or adding control loops, can become a tedious task, however.
To simplify this potentially error-prone process, a modular modeling approach is being proposed in this article.
It consists of small state-space models for certain building blocks of a converter system on the one hand, and standardized operations for connecting these subsystem models to an overall converter system model on the other hand.
The resulting state-space system model builds upon a two-port converter description and allows the extraction of control-oriented and impedance characteristics at any modeling stage, be it open loop or closed loop, single converter or series connections of converters.
The ease of creating more complex models enabled by the proposed approach is also demonstrated with examples comprising multiple control loops or cascaded converters.
}

\hypersetup{%
    pdfauthor={\AuthorRunning},
    pdftitle={\TitleRunning},
    pdfcreator={},pdfproducer={}
}

\begin{document}

\AddToShipoutPicture*{\AtPageUpperLeft{\put(\LenToUnit{2cm},\LenToUnit{-1.6cm}){%
   \color{black!50}%
   \begin{minipage}[c]{17cm}
   \footnotesize%
   This is a preprint version.
   The final article is available at \textcolor{blue!60}{\href{http://dx.doi.org/10.3390/j2030018}{doi:10.3390/j2030018}}.
   Please cite as:\\
   G. Herbst, ``A Building-Block Approach to State-Space Modeling of DC-DC Converter Systems,'' \emph{J}, vol. 2, no. 3, pp. 247--267, Jul. 2019.
   \end{minipage}
}}}

\maketitle

% -----------------------------------------------------------------------------
% Introduction
% -----------------------------------------------------------------------------

\section{Introduction}

Small-signal modeling of DC-DC converters has been a subject of research for several decades now, resulting in two main families of modeling techniques: state-space averaging \cite{Middlebrook:1976} and switch or circuit averaging \cite{Vorperian:1990,Czarkowski:1993}.
State-space averaging has become a broadly used modeling approach for DC-DC converters \cite{Frances:2018}, and is considered the most effective method for building small-signal models \cite{Zhan:2014}.
It~continuously received extensions, for example, for converters in discontinuous conduction mode (DCM)~\cite{Davoudi:2006,Sun:2001} or variable-frequency operation \cite{Priewasser:2014}, and new theoretical results to this day \cite{Meo:2018}.

The typical workflow using state-space averaging (SSA) includes applying Laplace transform and deriving frequency-domain transfer functions characterizing input/output impedance or control\hyp{}oriented behavior \cite{Petrovic:2004}, which must then be maintained individually.
Thus, whenever the converter is being extended, for example, by control loops, this leads to the tedious and potentially error-prone work of having to adapt all frequency-domain models separately \cite{Vesti:2013}.
Even worse, the underlying state-space model might have to be derived again if circuit elements (such as an input filter) are added to the converter. Basso \cite{Basso:2014} calls this the ``pain of the SSA technique.''

For easier modeling, the idea of modular approaches has emerged in the last two decades, be it for single converters \cite{Wu:1998}, or even working towards automatic modeling solutions for microgrids~\cite{Frances:2016,Frances:2017}.
Building and analyzing models for multi-converter configurations has of course been of increasing interest since the advent of DC microgrids \cite{Frances:2018,Haghmaram:2017}, but has been the subject of research long before, especially when studying converter interactions in a cascaded source-load setup \cite{Haroun:2014,Li:2015:ECCE,Mummadi:2009}.
Matters of particular interest are impedance analysis \cite{Ahmadi:2014,Ali:2019,Vesti:2013}, implications for controller design \cite{Ahmadi:2012,Haroun:2014,Pidaparthy:2015} and, of course, stability \cite{Ahmadi:2011,Li:2015:ECCE,Mummadi:2009}.
It has been pointed out \cite{Cupelli:2015,Tehrani:2018} that studying detailed dynamic interactions between source and load converters is important for stability analysis, going beyond constant power load~assumptions.

Many existing studies adopt an inverse-hybrid two-port description of converters (also known as g-parameters) \cite{Maranesi:1988}, which lends itself well to the behavioral modeling of converters and converter systems~\cite{Suntio:2014,Zenger:2006}, especially in the case of unterminated converter models \cite{Arnedo:2009,Cvetkovic:2013,Suntio:2018}.

Departing from the prevalent approach of deriving individual frequency-domain transfer functions for the different aspects of converter (system) modeling, this article proposes creating and maintaining a single state-space model in the time domain.
It encompasses both control-oriented and electrical characteristics in a two-port description with inverse-hybrid parameters (g-parameters), following existing good practice and enhanced by additional control inputs as necessary.
Whenever needed, of course, all relevant frequency-domain transfer functions can easily be extracted at any modeling stage from this model, superseding the necessity of updating individual frequency-domain transfer functions.

In order to relieve the ``SSA pain'' when building larger models, standardized (and therefore less error-prone) operations are being proposed for attaching controllers, closing control loops and~creating series-connections of converter subsystems.
These subsystems can be passive circuits such as input filters, open-loop or controlled converters, or even series-connections of converters themselves.
In that manner, the actual modeling effort is reduced to modeling smaller building blocks which can later on be reused and combined to create larger converter system models.

The remainder of this article is organized as follows:
The general modeling approach is presented in Section \ref{sec:General}.
Modeling examples for building blocks such as passive components, converters and~controllers are given in the Appendixes \ref{sec:Passive}--\ref{sec:Controller}, respectively.
From these ingredients, a complete model of a converter system can be built up incrementally.
For this purpose, new model connection operations are introduced in Section \ref{sec:ConverterController} (adding control loops to converters) and Section \ref{sec:Connecting} (series-connecting converter subsystems), which form the main contribution of this article.
Finally, examples for modeling a multiloop-controlled converter and a two-stage converter system are presented in Section \ref{sec:Examples}.

% -----------------------------------------------------------------------------
% Modular Modeling Approach
% -----------------------------------------------------------------------------

\section{Modular Modeling Approach}
\label{sec:General}
\vspace{-6pt}

% -----------------------------------------------------------------------------

\subsection{Model for Passive Components}
\label{sec:General_Passive}

For passive components of a converter system, such as an input filter or an electrical load, the~two-port state-space model is given in general form in \refEq{eqn:TwoPort_Passive}.
The number of state variables, and therefore the size of the matrices $\Matrix{A}$, $\Matrix{B}$, $\Matrix{C}$ depends on the order of the respective system, that is, the number of storage elements.

\begin{equation}
\label{eqn:TwoPort_Passive}
\begin{split}
\Vector{\dot{x}}(t)
&=
\Matrix{A}
\cdot
\Vector{x}(t)
+
\begin{pmatrix}
\Matrix{B}_1  &  \Matrix{B}_2
\end{pmatrix}
\cdot
\begin{pmatrix}
v_\mathrm{in}(t)
\\
i_\mathrm{out}(t)
\end{pmatrix}
\\
\begin{pmatrix}
i_\mathrm{in}(t)
\\
v_\mathrm{out}(t)
\end{pmatrix}
&=
\begin{pmatrix}
\Matrix{C}_1
\\
\Matrix{C}_2
\end{pmatrix}
\cdot
\Vector{x}(t)
+
\begin{pmatrix}
D_{11}  &  D_{12}
\\
D_{21}  &  D_{22}
\end{pmatrix}
\cdot
\begin{pmatrix}
v_\mathrm{in}(t)
\\
i_\mathrm{out}(t)
\end{pmatrix}
\raisetag{8ex}
\end{split}
\end{equation}

\begin{figure}[h]
    \centering%
    \includegraphics{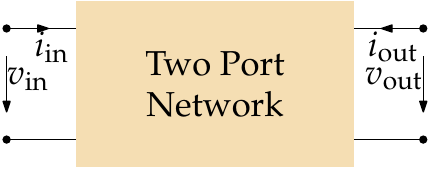}%
    \caption{Generic two-port network. Currents are defined as flowing into the input and output ports.}
    \label{fig:TwoPort_General}
\end{figure}

\refFigBegin{fig:TwoPort_General} illustrates the definition of input and output voltages and currents for a generic two-port system.
Examples for modeling passive subsystems of a converter system are given in Appendix \ref{sec:Passive}.

% -----------------------------------------------------------------------------

\subsection{Model for Controlled Converters and Converter Systems}
\label{sec:General_Converter}

Based on Section \ref{sec:General_Passive}, the model for a converter (system) gains at least one additional input: the~control signal $\mathit{ctl}(t)$.
For a single DC-DC converter, this will be the duty cycle in most cases.
\refEqBegin{eqn:TwoPort_Passive} is therefore enhanced to \refEq{eqn:TwoPort_Controlled}.
Examples are given or referenced in Appendix \ref{sec:Converter}.

\begin{equation}
\label{eqn:TwoPort_Controlled}
\begin{split}
\Vector{\dot{x}}(t)
&=
\Matrix{A}
\cdot
\Vector{x}(t)
+
\begin{pmatrix}
\Matrix{B}_1  &  \Matrix{B}_2  &  \Matrix{B}_3
\end{pmatrix}
\cdot
\begin{pmatrix}
v_\mathrm{in}(t)
\\
i_\mathrm{out}(t)
\\
\mathit{ctl}(t)
\end{pmatrix}
\\
\begin{pmatrix}
i_\mathrm{in}(t)
\\
v_\mathrm{out}(t)
\end{pmatrix}
&=
\begin{pmatrix}
\Matrix{C}_1
\\
\Matrix{C}_2
\end{pmatrix}
\cdot
\Vector{x}
+
\begin{pmatrix}
D_{11}  &  D_{12}  &  D_{13}
\\
D_{21}  &  D_{22}  &  D_{23}
\end{pmatrix}
\cdot
\begin{pmatrix}
v_\mathrm{in}(t)
\\
i_\mathrm{out}(t)
\\
\mathit{ctl}(t)
\end{pmatrix}
\raisetag{12ex}
\end{split}
\end{equation}

\refEqBegin{eqn:TwoPort_Controlled} is a very general model.
Depending on both the approach for controlling the converter (such as voltage mode or current mode) as well as the modeling stage during the elaboration of the model, there are different possible meanings or functions the control signal $\mathit{ctl}(t)$ can possess: duty ratio, peak or average current reference (in current mode control, if the model only describes the current loop so far), voltage reference (in voltage mode or if the outer voltage control loop around a current-controlled converter has been closed), or even the respective open-loop inputs if the control loop has not yet been closed in the model.

Additionally, \refEq{eqn:TwoPort_Controlled} is very general in the regard that the model may contain a (series) connection of several controlled converters.
In this case, $\mathit{ctl}(t)$ is a vector containing the individual control loop inputs for the converters in this model.

When connecting input filters or loads to the model according to \refEq{eqn:TwoPort_Controlled}, its structure remains the same, therefore \refEq{eqn:TwoPort_Controlled} is being used both for modeling ``bare'' converters as well as the final result, that is, a~converter system with one or more controlled converters as well as possible filter and load subsystems.

% -----------------------------------------------------------------------------

\subsection{Model for Controllers}
\label{sec:General_Controller}

While controllers---such as a controller for the output voltage or an underlying average current mode controller---will be modeled in this article in a state-space representation as well, they require a different treatment and do not fit into the two-port model category.
This is due to the fact that even though the controllers may be realized in analog form by means of electric circuits, they represent an information flow rather than a power flow.
The controller model employed here will be a continuous-time state-space model given in \refEq{eqn:General_Controller}, with a single input (the control error value $e$) and a single output (the control signal $u$ of the respective controller).
Examples of controllers typically used for converters are given in Appendix \ref{sec:Controller}.

\begin{equation}
\label{eqn:General_Controller}
\begin{split}
\Vector{\dot{x}}_\mathrm{C}(t) &= \Matrix{A}_\mathrm{C} \cdot \Vector{x}_\mathrm{C}(t) + \Matrix{B}_\mathrm{C} \cdot e(t)
\\
u(t) &= \Matrix{C}_\mathrm{C} \cdot \Vector{x}_\mathrm{C}(t) + \Matrix{D}_\mathrm{C} \cdot e(t)
\end{split}
\end{equation}

% -----------------------------------------------------------------------------

\subsection{Making Connections}
\label{sec:General_Connections}

\begin{figure}[h]
    \centering%
    \includegraphics{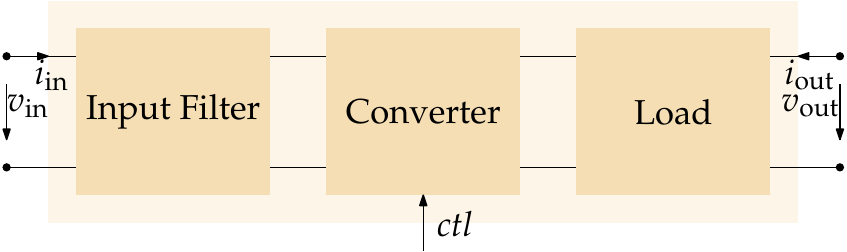}%
    \caption{Two-port network description consisting of three subsystems: input filter, converter (with a control input $\mathit{ctl}$), and the actual load.}
    \label{fig:TwoPort_FilterConverterLoad}
\end{figure}

The most simple form of a practical converter system consists of the actual ``bare'' converter (such~as a buck or boost converter), an (optional) input filter and a load, cf.\ \refFig{fig:TwoPort_FilterConverterLoad}.
A more complicated case arises when a second converter is attached as the load system of a first converter.

In this article, a solution is presented to create a complete system model for arbitrary series connections of converter subsystems.
The generic model ingredients have already been presented in Section \ref{sec:General_Passive} (passive components), Section \ref{sec:General_Converter} (converters), and Section \ref{sec:General_Controller} (controllers).
Two~different methods of connecting these model ingredients to a complete system mode are necessary: (1)~connecting a converter with one or more controllers, which will be presented in Section \ref{sec:ConverterController}, and (2)~connecting (controlled) converter subsystems or passive subsystems in a series connection, which will be presented in Section \ref{sec:Connecting}.

A typical workflow of using the models and methods presented in this article is as follows: (1)~create a converter model (examples in Appendix \ref{sec:Converter}), (2) create a controller model (examples in Appendix \ref{sec:Controller}) and connect the controller (Section \ref{sec:ConverterController}) for every control loop of the converter, (3) make series connections to further controlled converters (if present, Section \ref{sec:Connecting}), (4) create models for input filter and load (examples in Appendix \ref{sec:Passive}) and connect them (Section \ref{sec:Connecting}).
The latter may, of course, also be connected before connecting the controller in order to tune the controller according to possible interactions of the converter with source and load impedances, especially if the system consists only of a single converter.

% -----------------------------------------------------------------------------
% Adding Control Loops to Converter Models
% -----------------------------------------------------------------------------

\section{Adding Control Loops to Converter Models}
\label{sec:ConverterController}

In the following, a general procedure is being described that combines a two-port converter model featuring one (scalar) control input $\mathit{ctl}(t)$ as given in \refEq{eqn:TwoPort_Controlled} with a model of the controller according to \refEq{eqn:General_Controller}.
Firstly, an open-loop model is being constructed, followed by the closed-loop model, both of which maintaining the same interface of the general two-port state-space model \refEq{eqn:TwoPort_Controlled}.

\begin{figure}[h]
    \centering%
    \subfloat[]{%
        \includegraphics{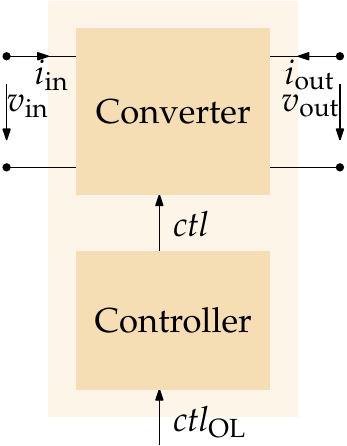}%
        \label{fig:ConverterController_InterfaceOL}%
    }
    \hspace{2cm}%
    \subfloat[]{%
        \includegraphics{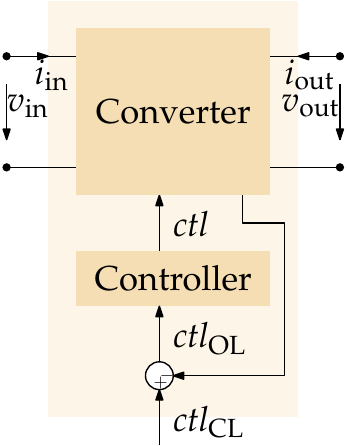}%
        \label{fig:ConverterController_InterfaceCL}%
    }
    \caption{Interfaces of the two-port network models with control inputs when adding a controller to a converter. {(\textbf{a})} Open-loop case. {(\textbf{b})} Closed-loop case.}
    \label{fig:ConverterController_Interface}
\end{figure}

The control input $\mathit{ctl}(t)$ of the given converter could be the duty cycle or the set-point of an inner control loop, such as an inner current control loop.
Through the stages of connecting the converter model with the controller model, the meaning of the model's control input changes from its original meaning to the control error value $e(t)$ (open-loop case) and finally the set-point (reference signal) $r(t)$ of the closed control loop.
The interfaces of the respective two-port models are shown in \refFig{fig:ConverterController_Interface}.

In order to obtain the open-loop model, the state variable vector of the converter model must be enhanced by the state variables of the controller model.
In general, a common model structure of the open-loop and closed-loop models can be given as follows in \refEq{eqn:ConverterController_General}, with subscripts OL and CL distinguishing the open-loop and closed-loop case.

\begin{equation}
\label{eqn:ConverterController_General}
\begin{split}
\begin{pmatrix}
\Vector{\dot{x}}(t)
\\
\Vector{\dot{x}}_\mathrm{C}(t)
\end{pmatrix}
&=
\Matrix{A}_\mathrm{OL/CL}
\cdot
\begin{pmatrix}
\Vector{x}(t)
\\
\Vector{x}_\mathrm{C}(t)
\end{pmatrix}
+
\Matrix{B}_\mathrm{OL/CL}
\cdot
\begin{pmatrix}
v_\mathrm{in}(t)
\\
i_\mathrm{out}(t)
\\
\mathit{ctl}_\mathrm{OL/CL}(t)
\end{pmatrix}
\\
\begin{pmatrix}
i_\mathrm{in}(t)
\\
v_\mathrm{out}(t)
\end{pmatrix}
&=
\Matrix{C}_\mathrm{OL/CL}
\cdot
\begin{pmatrix}
\Vector{x}(t)
\\
\Vector{x}_\mathrm{C}(t)
\end{pmatrix}
+
\Matrix{D}_\mathrm{OL/CL}
\cdot
\begin{pmatrix}
v_\mathrm{in}(t)
\\
i_\mathrm{out}(t)
\\
\mathit{ctl}_\mathrm{OL/CL}(t)
\end{pmatrix}
\\
\text{with} \quad {ctl}_\mathrm{OL/CL}(t) &=
\begin{cases}
{ctl}_\mathrm{OL}(t) = e(t) & \quad \text{open-loop case (control error)}
\\
{ctl}_\mathrm{CL}(t) = r(t) & \quad \text{closed-loop case (reference signal)}
\end{cases}
\end{split}
\end{equation}

The main step required to obtain the $\Matrix{A}$, $\Matrix{B}$, $\Matrix{C}$, $\Matrix{D}$ matrices of the open-loop model is to account for the changed meaning of the control input of the model.
The (now internal) control input of the converter must be connected to the output of the controller, cf.\ \refEq{eqn:ConverterController_OpenLoop}.

\begin{equation}
\begin{aligned}
\label{eqn:ConverterController_OpenLoop}
&\Matrix{A}_\mathrm{OL} =
\begin{pmatrix}
\Matrix{A}  &  \Matrix{B}_3 \cdot \Matrix{C}_\mathrm{C}
\\
\Matrix{0}^{m \times n}  &  \Matrix{A}_\mathrm{C}
\end{pmatrix}
,
& \quad &\Matrix{B}_\mathrm{OL} =
\begin{pmatrix}
\Matrix{B}_1  &  \Matrix{B}_2  &  \Matrix{B}_3 \cdot D_\mathrm{C}
\\
\Matrix{0}^{m \times 1}  &  \Matrix{0}^{m \times 1}  &  \Matrix{B}_\mathrm{C}
\end{pmatrix}
,
\\
&\Matrix{C}_\mathrm{OL} =
\begin{pmatrix}
\Matrix{C}_1  &  \Matrix{0}^{1 \times m}
\\
\Matrix{C}_2  &  \Matrix{0}^{1 \times m}
\end{pmatrix}
,
& \quad &\Matrix{D}_\mathrm{OL} =
\begin{pmatrix}
D_{11}  &  D_{12}  &  D_{13} \cdot D_\mathrm{C}
\\
D_{21}  &  D_{22}  &  D_{23} \cdot D_\mathrm{C}
\end{pmatrix}
\end{aligned}
\end{equation}
In \refEq{eqn:ConverterController_OpenLoop}, $n$ refers to the number of state variables in the converter model, and $m$ to the number of state variables in the controller model.

When deriving the closed-loop model in the following, it will be assumed that there is no (or only a negligible) direct feedthrough from the original control input to the desired control variable.
If, for example, the output voltage is the desired control variable, $D_{23} \approx 0$ should hold, which is the case for all converter types mentioned in this article.

The $\Matrix{B}$, $\Matrix{C}$, $\Matrix{D}$ matrices of the open-loop and closed-loop case are identical, that is, $\Matrix{B}_\mathrm{CL} = \Matrix{B}_\mathrm{OL}$, $\Matrix{C}_\mathrm{CL} = \Matrix{C}_\mathrm{OL}$, $\Matrix{D}_\mathrm{CL} = \Matrix{D}_\mathrm{OL}$.
The $\Matrix{A}$ matrix of the closed-loop model is computed via \refEq{eqn:ConverterController_ClosedLoop_A}, which is an equation known from classical state feedback control.

\begin{equation}
\label{eqn:ConverterController_ClosedLoop_A}
\Matrix{A}_\mathrm{CL} = \Matrix{A}_\mathrm{OL} - \Matrix{B}_\mathrm{OL} \cdot \Matrix{K}
\end{equation}

A suitable feedback gain matrix $\Matrix{K}$ must be chosen depending on which control loop is being closed, such as an inner average current loop or an output voltage loop.
It is important to note that $\Matrix{K}$ is not being used to design the controller (as in state feedback control), but only to select and feed back the desired control variable when constructing the closed-loop state-space model.
The~controller parameters are contained in the controller model \refEq{eqn:General_Controller}.
To demonstrate that the choice of $\Matrix{K}$ is almost trivial in most applications, the (probably) most important two cases will be shown in Sections \ref{sec:ConverterController_CurrentMode} and \ref{sec:ConverterController_VoltageMode}.

As a final side note, a sensor or filter with non-negligible dynamics in the feedback path can rather easily be accounted for by extending the converter model with the respective dynamics and choosing the output of the sensor or filter as the control variable.

% -----------------------------------------------------------------------------

\subsection{Average Current Mode}
\label{sec:ConverterController_CurrentMode}
For closing a current loop, $\Matrix{K}$ must be chosen in order to extract the desired current variable from the vector of converter system states $\Vector{x}(t)$ and feed it back to the third input of the open-loop converter model, that is, $\mathit{ctl}_\mathrm{OL}(t)$.
If we, for example, assume for the converter model that the average current is the first state variable (as in the boost converter example in Appendix \ref{sec:Converter_Example_Boost}), $\Matrix{K}$ simply has to be chosen as~\refEq{eqn:K_CurrentMode}.

\begin{equation}
\label{eqn:K_CurrentMode}
\Matrix{K} =
\begin{pmatrix}
\Matrix{0}^{2 \times 1}  &  \Matrix{0}^{2 \times (n-1+m)}
\\
1  &  \Matrix{0}^{1 \times (n-1+m)}
\end{pmatrix}
\end{equation}

% -----------------------------------------------------------------------------

\subsection{Voltage Mode or Outer Voltage Loops}
\label{sec:ConverterController_VoltageMode}
If output voltage $v_\mathrm{out}(t)$ is the desired control variable, which is the case for voltage mode as well as for an outer voltage loop around a current-controlled converter, $\Matrix{K}$ has to be chosen as \refEq{eqn:K_VoltageMode}.

\begin{equation}
\label{eqn:K_VoltageMode}
\Matrix{K} =
\begin{pmatrix}
\Matrix{0}^{2 \times n}  &  \Matrix{0}^{2 \times m}
\\
\Matrix{C}_2  &  \Matrix{0}^{1 \times m}
\end{pmatrix}
\end{equation}
This assumes that $D_{23} \approx 0$ holds for the uncontrolled converter model, which means that $v_\mathrm{out}(t)$ can approximately be extracted from the state variable vector by $v_\mathrm{out}(t) \approx \Matrix{C}_2 \cdot \Vector{x}(t)$.

% -----------------------------------------------------------------------------
% Connecting Converter Models
% -----------------------------------------------------------------------------

\section{Connecting Converter Models}
\label{sec:Connecting}

In order to create a combined state-space model for a series connection of two subsystems (denoted ``source'' and ``load'' here), the respective models for source (superscript S) and load (superscript L) must be given in the following form:

\begin{equation}
\label{eqn:ConnectSystems_SL}
\begin{split}
\Vector{\dot{x}}^\SL(t)
&=
\Matrix{A}^\SL
\cdot
\Vector{x}^\SL(t)
+
\Matrix{B}^\SL
\cdot
\begin{pmatrix}
v_\mathrm{in}^\SL(t)  \\[1ex]  i_\mathrm{out}^\SL(t)  \\[1ex]  \Vector{\mathit{ctl}}^\SL(t)
\end{pmatrix}
\\
\begin{pmatrix}
i_\mathrm{in}^\SL(t)  \\[1ex]  v_\mathrm{out}^\SL(t)
\end{pmatrix}
&=
\Matrix{C}^\SL
\cdot
\Vector{x}^\SL(t)
+
\Matrix{D}^\SL
\cdot
\begin{pmatrix}
v_\mathrm{in}^\SL(t)  \\[1ex]  i_\mathrm{out}^\SL(t)  \\[1ex]  \Vector{\mathit{ctl}}^\SL(t)
\end{pmatrix}
\end{split}
\end{equation}

Note that the connection operation introduced in this section works regardless of the number of control inputs $\Vector{\mathit{ctl}}(t)$ in the source and load models, that is, even without control inputs, as is the case for passive subsystems such as filters or loads. In these cases, the respective columns of the $\Matrix{B}$ and $\Matrix{D}$ matrices must simply be left out in all of the following equations.

\begin{figure}[h]
    \centering%
    \includegraphics{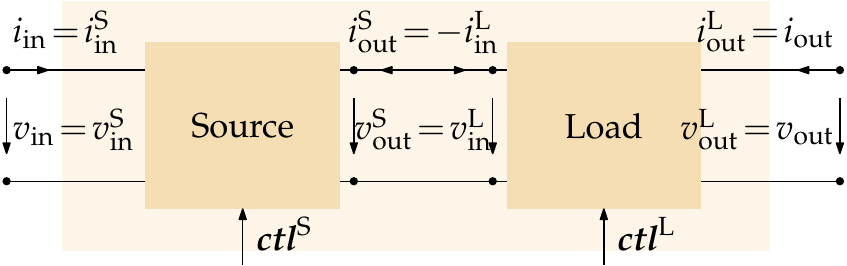}%
    \caption{Series connection of two two-port networks (source and load). The control inputs $\Vector{\mathit{ctl}}^\SL(t)$ are optional and only present if source or load contain at least one converter, respectively.}
    \label{fig:TwoPort_SourceLoad}
\end{figure}

The input and output ports of the combined system are the input port of the source subsystem and the output port of the load subsystem, that is, the following equations hold:
$i_\mathrm{in}(t) = i_\mathrm{in}^\S(t)$,
$v_\mathrm{in}(t) = v_\mathrm{in}^\S(t)$,
$v_\mathrm{out}(t) = v_\mathrm{out}^\L(t)$, and
$i_\mathrm{out}(t) = i_\mathrm{out}^\L(t)$,
cf.\ \refFig{fig:TwoPort_SourceLoad}.
The result is the following combined model \refEq{eqn:ConnectSystems_Model}.

\begin{equation}
\label{eqn:ConnectSystems_Model}
\begin{split}
\begin{pmatrix}
\Vector{\dot{x}}^\S(t)  \\  \Vector{\dot{x}}^\L(t)
\end{pmatrix}
&=
\Matrix{A}
\cdot
\begin{pmatrix}
\Vector{x}^\S(t)  \\  \Vector{x}^\L(t)
\end{pmatrix}
+
\Matrix{B}
\cdot
\begin{pmatrix}
v_\mathrm{in}(t)  \\  i_\mathrm{out}(t)  \\  \Vector{\mathit{ctl}}^\S(t)  \\  \Vector{\mathit{ctl}}^\L(t)
\end{pmatrix}
\\
\begin{pmatrix}
i_\mathrm{in}(t)  \\  v_\mathrm{out}(t)
\end{pmatrix}
&=
\Matrix{C}
\cdot
\begin{pmatrix}
\Vector{x}^\S(t)  \\  \Vector{x}^\L(t)
\end{pmatrix}
+
\Matrix{D}
\cdot
\begin{pmatrix}
v_\mathrm{in}(t)  \\  i_\mathrm{out}(t)  \\  \Vector{\mathit{ctl}}^\S(t)  \\  \Vector{\mathit{ctl}}^\L(t)
\end{pmatrix}
\end{split}
\end{equation}

The equations of the resulting model can be found by replacing the inner terminal variables of the two-port series connection, which are not present in \refEq{eqn:ConnectSystems_Model} anymore, cf.\ \refFig{fig:TwoPort_SourceLoad}.
For the inner voltage $v_\mathrm{out}^\S(t) = v_\mathrm{in}^\L(t)$ and the inner current $i_\mathrm{out}^\S(t) = -i_\mathrm{in}^\L(t)$ the following equations hold:

\begin{align}
v_\mathrm{in}^\L = v_\mathrm{out}^\S
&= \Matrix{C}_2^\S \Vector{x}^\S + D_{21}^\S v_\mathrm{in}^\S + D_{22}^\S i_\mathrm{out}^\S + \Matrix{D}_{23}^\S \Vector{\mathit{ctl}}^\S
\notag\\
&= \Matrix{C}_2^\S \Vector{x}^\S + D_{21}^\S v_\mathrm{in}^\S + D_{22}^\S \left( -\Matrix{C}_1^\L \Vector{x}^\L - D_{11}^\L v_\mathrm{in}^\L - D_{12}^\L i_\mathrm{out}^\L - \Matrix{D}_{13}^\L \Vector{\mathit{ctl}}^\L \right) + \Matrix{D}_{23}^\S \Vector{\mathit{ctl}}^\S
\notag\\
&= \frac{1}{ 1 + D_{11}^\L D_{22}^\S } \cdot \left( \Matrix{C}_2^\S \Vector{x}^\S - D_{22}^\S \Matrix{C}_1^\L \Vector{x}^\L + D_{21}^\S v_\mathrm{in} - D_{22}^\S D_{12}^\L i_\mathrm{out} + \Matrix{D}_{23}^\S \Vector{\mathit{ctl}}^\S - D_{22}^\S \Matrix{D}_{13}^\L \Vector{\mathit{ctl}}^\L \right)
\label{eqn:ConnectSystems_Replace_VInL}
\end{align}

\begin{align}
i_\mathrm{out}^\S = -i_\mathrm{in}^\L
&= -\Matrix{C}_1^\L \Vector{x}^\L - D_{11}^\L v_\mathrm{in}^\L - D_{12}^\L i_\mathrm{out}^\L - \Matrix{D}_{13}^\L \Vector{\mathit{ctl}}^\L
\notag\\
&= -\Matrix{C}_1^\L \Vector{x}^\L - D_{11}^\L \left( \Matrix{C}_2^\S \Vector{x}^\S + D_{21}^\S v_\mathrm{in}^\S + D_{22}^\S i_\mathrm{out}^\S + \Matrix{D}_{23}^\S \Vector{\mathit{ctl}}^\S \right) - D_{12}^\L i_\mathrm{out}^\L - \Matrix{D}_{13}^\L \Vector{\mathit{ctl}}^\L
\notag\\
&= \frac{-1}{ 1 + D_{11}^\L D_{22}^\S } \cdot \left( D_{11}^\L \Matrix{C}_2^\S \Vector{x}^\S + \Matrix{C}_1^\L \Vector{x}^\L + D_{11}^\L D_{21}^\S v_\mathrm{in} + D_{12}^\L i_\mathrm{out} + D_{11}^\L \Matrix{D}_{23}^\S \Vector{\mathit{ctl}}^\S + \Matrix{D}_{13}^\L \Vector{\mathit{ctl}}^\L \right)
\label{eqn:ConnectSystems_Replace_IOutS}
\end{align}

Putting \refEq{eqn:ConnectSystems_Replace_VInL} and \refEq{eqn:ConnectSystems_Replace_IOutS} in the source/load state equations of \refEq{eqn:ConnectSystems_SL} will now lead to the combined state equations in \refEq{eqn:ConnectSystems_Model}.
Similarly, the output equation of the connected system in \refEq{eqn:ConnectSystems_Model} is obtained by putting \refEq{eqn:ConnectSystems_Replace_VInL} and \refEq{eqn:ConnectSystems_Replace_IOutS} in the output equations from \refEq{eqn:ConnectSystems_SL} for $i_\mathrm{in}(t) = i_\mathrm{in}^\S(t)$ and $v_\mathrm{out}(t) = v_\mathrm{out}^\L(t)$.
The resulting detailed formulation of the matrix contents for $\Matrix{A}$, $\Matrix{B}$, $\Matrix{C}$, $\Matrix{D}$ in \refEq{eqn:ConnectSystems_Model} is given in \refEq{eqn:ConnectSystems_ABCD}.

\begin{equation}
\label{eqn:ConnectSystems_ABCD}
\begin{split}
\Matrix{A} &=
\begin{pmatrix}
\D \Matrix{A}^\S - \frac{ \Matrix{B}_2^\S D_{11}^\L \Matrix{C}_2^\S }{ 1 + D_{11}^\L D_{22}^\S }
&
\D -\frac{ \Matrix{B}_2^\S \Matrix{C}_1^\L }{ 1 + D_{11}^\L D_{22}^\S }
\\[1.5em]
\D \frac{ \Matrix{B}_1^\L \Matrix{C}_2^\S }{ 1 + D_{11}^\L D_{22}^\S }
&
\D \Matrix{A}^\L - \frac{ \Matrix{B}_1^\L D_{22}^\S \Matrix{C}_1^\L }{ 1 + D_{11}^\L D_{22}^\S }
\end{pmatrix}
,
\quad
\Matrix{C} =
\begin{pmatrix}
\D \Matrix{C}_1^\S - \frac{ D_{12}^\S D_{11}^\L \Matrix{C}_2^\S }{ 1 + D_{11}^\L D_{22}^\S }
&
\D -\frac{ D_{12}^\S \Matrix{C}_1^\L }{ 1 + D_{11}^\L D_{22}^\S }
\\[1.5em]
\D \frac{ D_{21}^\L \Matrix{C}_2^\S }{ 1 + D_{11}^\L D_{22}^\S }
&
\D \Matrix{C}_2^\L - \frac{ D_{21}^\L D_{22}^\S \Matrix{C}_1^\L }{ 1 + D_{11}^\L D_{22}^\S }
\end{pmatrix}
,
\\
\Matrix{B} &=
\begin{pmatrix}
\D \Matrix{B}_1^\S - \frac{ \Matrix{B}_2^\S D_{11}^\L D_{21}^\S }{ 1 + D_{11}^\L D_{22}^\S }
&
\D -\frac{ \Matrix{B}_2^\S D_{12}^\L }{ 1 + D_{11}^\L D_{22}^\S }
&
\D \Matrix{B}_3^\S - \frac{ \Matrix{B}_2^\S D_{11}^\L \Matrix{D}_{23}^\S }{ 1 + D_{11}^\L D_{22}^\S }
&
\D -\frac{ \Matrix{B}_2^\S \Matrix{D}_{13}^\L }{ 1 + D_{11}^\L D_{22}^\S }
\\[1.5em]
\D \frac{ \Matrix{B}_1^\L D_{21}^\S }{ 1 + D_{11}^\L D_{22}^\S }
&
\D \Matrix{B}_2^\L - \frac{ \Matrix{B}_1^\L D_{22}^\S D_{12}^\L }{ 1 + D_{11}^\L D_{22}^\S }
&
\D \frac{ \Matrix{B}_1^\L \Matrix{D}_{23}^\S }{ 1 + D_{11}^\L D_{22}^\S }
&
\D \Matrix{B}_3^\L - \frac{ \Matrix{B}_1^\L D_{22}^\S \Matrix{D}_{13}^\L }{ 1 + D_{11}^\L D_{22}^\S }
\end{pmatrix}
,
\\
\Matrix{D} &=
\begin{pmatrix}
\D D_{11}^\S - \frac{ D_{12}^\S D_{11}^\L D_{21}^\S }{ 1 + D_{11}^\L D_{22}^\S }
&
\D -\frac{ D_{12}^\S D_{12}^\L }{ 1 + D_{11}^\L D_{22}^\S }
&
\D \Matrix{D}_{13}^\S - \frac{ D_{12}^\S D_{11}^\L \Matrix{D}_{23}^\S }{ 1 + D_{11}^\L D_{22}^\S }
&
\D -\frac{ D_{12}^\S \Matrix{D}_{13}^\L }{ 1 + D_{11}^\L D_{22}^\S }
\\[1.5em]
\D \frac{ D_{21}^\L D_{21}^\S }{ 1 + D_{11}^\L D_{22}^\S }
&
\D D_{22}^\L - \frac{ D_{21}^\L D_{22}^\S D_{12}^\L }{ 1 + D_{11}^\L D_{22}^\S }
&
\D \frac{ D_{21}^\L \Matrix{D}_{23}^\S }{ 1 + D_{11}^\L D_{22}^\S }
&
\D \Matrix{D}_{23}^\L - \frac{ D_{21}^\L D_{22}^\S \Matrix{D}_{13}^\L }{ 1 + D_{11}^\L D_{22}^\S }
\end{pmatrix}
\end{split}
\end{equation}

It has to be noted that \refEq{eqn:ConnectSystems_ABCD} will often be reduced to a simpler form in practical cases since not all matrix elements of the subsystem matrices are non-zero.
Furthermore, \refEq{eqn:ConnectSystems_Model} and \refEq{eqn:ConnectSystems_ABCD} are primarily meant to be implemented in a computer algebra system or a numerical software package only once and then be used to conveniently build up a complete model of a DC-DC converter system from simple subsystems in a fully or semi-automated manner.
Compact matrix formulations for \refEq{eqn:ConnectSystems_ABCD} can be found by inspection, the corresponding set of equations is presented in \refEq{eqn:ConnectSystems_ABCD_Matrix}.

\begin{equation}
\label{eqn:ConnectSystems_ABCD_Matrix}
\begin{split}
\Matrix{A} &=
\diag\left( \Matrix{A}^\S, \Matrix{A}^\L \right) - \Matrix{M}_1 \cdot \Matrix{M}_C
\\
\Matrix{B} &=
\left( \begin{array}{c|c}
\diag\left( \Matrix{B}_1^\S, \Matrix{B}_2^\L \right)  &   \diag\left( \Matrix{B}_3^\S, \Matrix{B}_3^\L \right)
\end{array} \right)
- \Matrix{M}_1 \cdot \Matrix{M}_D
\\
\Matrix{C} &=
\diag\left( \Matrix{C}_1^\S, \Matrix{C}_2^\L \right) - \Matrix{M}_2 \cdot \Matrix{M}_C
\\
\Matrix{D} &=
\left( \begin{array}{c|c}
\diag\left( D_{11}^\S, D_{22}^\L \right)  &   \diag\left( \Matrix{D}_{13}^\S, \Matrix{D}_{23}^\L \right)
\end{array} \right)
- \Matrix{M}_2 \cdot \Matrix{M}_D
\end{split}
\raisetag{7ex}
\end{equation}

In \refEq{eqn:ConnectSystems_ABCD_Matrix}, the abbreviations defined in \refEq{eqn:ConnectSystems_ABCD_Matrix_Helper} were used for recurring terms.

\begin{equation}
\label{eqn:ConnectSystems_ABCD_Matrix_Helper}
\begin{aligned}
\Matrix{M}_1 &= \diag\left( \Matrix{B}_2^\S, \Matrix{B}_1^\L \right)
\cdot \begin{pmatrix}
D_{22}^\S  &  -1
\\
1  &  D_{11}^1 \L
\end{pmatrix}^{-1}
\\
\Matrix{M}_2 &= \diag\left( D_{12}^\S, D_{21}^\L \right)
\cdot \begin{pmatrix}
D_{22}^\S  &  -1
\\
1  &  D_{11}^\L
\end{pmatrix}^{-1}
\\
\Matrix{M}_C &= \diag\left( \Matrix{C}_2^\S, \Matrix{C}_1^\L \right)
\\
\Matrix{M}_D &= \left( \begin{array}{c|c}
\diag\left( D_{21}^\S, D_{12}^\L \right)  &   \diag\left( \Matrix{D}_{23}^\S, \Matrix{D}_{13}^\L \right)
\end{array} \right)
\\
\end{aligned}
\end{equation}

To summarize the connection procedure, all that is required to obtain the connected model from two given source and load models is to arrange the input, output, and state variables of the resulting system as given in \refEq{eqn:ConnectSystems_Model}, and compute the  $\Matrix{A}$, $\Matrix{B}$, $\Matrix{C}$, $\Matrix{D}$ matrices using \refEq{eqn:ConnectSystems_ABCD}.

% -----------------------------------------------------------------------------
% Summary
% -----------------------------------------------------------------------------

\section{Summary of the Building-Block Modeling Approach}
\label{sec:Summary}

For convenience, all building block types and connection operations of the modular modeling approach presented in this article are summarized in this section. Furthermore, the relation of the developed state-space model to well-known frequency-domain models of converter systems is given.

% -----------------------------------------------------------------------------

\subsection{Building Block Types}
\label{sec:BuildingBlocks}

To start with modeling a DC-DC converter or converter system, the first step is to collect all building blocks, such as models for the bare converter, controllers, filters, loads and so forth. In the modeling approach presented in this article, there are three different categories of building blocks in terms of their state-space model representation. For convenience, they are listed again in \refTable{tab:BuildingBlocks}. Concrete examples for these three building block types are given in the Appendixes \ref{sec:Passive}--\ref{sec:Controller}.

\begin{table}[h]
\caption{Summary of building block types in terms of their state space equation structure.}
\label{tab:BuildingBlocks}
\centering
\begin{tabular}{lll}
\toprule
\textbf{Type}  &  \textbf{Equation} &  \textbf{Definition}  \\
\midrule
%
% -------------------------------------
%
Passive components
&
(\ref{eqn:TwoPort_Passive})
&
\parbox{8cm}{
$\begin{aligned}
\Vector{\dot{x}}(t)
&=
\Matrix{A}
\cdot
\Vector{x}(t)
+
\begin{pmatrix}
\Matrix{B}_1  &  \Matrix{B}_2
\end{pmatrix}
\cdot
\begin{pmatrix}
v_\mathrm{in}(t)
\\
i_\mathrm{out}(t)
\end{pmatrix}
\\
\begin{pmatrix}
i_\mathrm{in}(t)
\\
v_\mathrm{out}(t)
\end{pmatrix}
&=
\begin{pmatrix}
\Matrix{C}_1
\\
\Matrix{C}_2
\end{pmatrix}
\cdot
\Vector{x}(t)
+
\begin{pmatrix}
D_{11}  &  D_{12}
\\
D_{21}  &  D_{22}
\end{pmatrix}
\cdot
\begin{pmatrix}
v_\mathrm{in}(t)
\\
i_\mathrm{out}(t)
\end{pmatrix}
\end{aligned}$}
\\ \addlinespace[\aboverulesep] \cmidrule{1-3} \addlinespace[\belowrulesep]
%
% -------------------------------------
%
Controlled converters
&
(\ref{eqn:TwoPort_Controlled})
&
\parbox{8cm}{
$\begin{aligned}
\Vector{\dot{x}}(t)
&=
\Matrix{A}
\cdot
\Vector{x}(t)
+
\begin{pmatrix}
\Matrix{B}_1  &  \Matrix{B}_2  &  \Matrix{B}_3
\end{pmatrix}
\cdot
\begin{pmatrix}
v_\mathrm{in}(t)
\\
i_\mathrm{out}(t)
\\
\mathit{ctl}(t)
\end{pmatrix}
\\
\begin{pmatrix}
i_\mathrm{in}(t)
\\
v_\mathrm{out}(t)
\end{pmatrix}
&=
\begin{pmatrix}
\Matrix{C}_1
\\
\Matrix{C}_2
\end{pmatrix}
\cdot
\Vector{x}
+
\begin{pmatrix}
D_{11}  &  D_{12}  &  D_{13}
\\
D_{21}  &  D_{22}  &  D_{23}
\end{pmatrix}
\cdot
\begin{pmatrix}
v_\mathrm{in}(t)
\\
i_\mathrm{out}(t)
\\
\mathit{ctl}(t)
\end{pmatrix}
\end{aligned}$}
\\ \addlinespace[\aboverulesep] \cmidrule{1-3} \addlinespace[\belowrulesep]
%
% -------------------------------------
%
Controllers
&
(\ref{eqn:General_Controller})
&
\parbox{8cm}{
$\begin{aligned}
\Vector{\dot{x}}_\mathrm{C}(t) &= \Matrix{A}_\mathrm{C} \cdot \Vector{x}_\mathrm{C}(t) + \Matrix{B}_\mathrm{C} \cdot e(t)
\\
u(t) &= \Matrix{C}_\mathrm{C} \cdot \Vector{x}_\mathrm{C}(t) + \Matrix{D}_\mathrm{C} \cdot e(t)
\end{aligned}$}
\\
\bottomrule
\end{tabular}
\end{table}

% -----------------------------------------------------------------------------

\subsection{Connecting Building Blocks}
\label{sec:Summary_Connecting}

There are two different kinds of connections that can be made in order to create a converter system model from smaller building blocks:
\begin{enumerate}
\item
\emph{Connecting converters with passive components or other converters.}
Since both building blocks for passive elements as given by \refEq{eqn:TwoPort_Passive} and building blocks for converters as given by \refEq{eqn:TwoPort_Controlled} are two-port models, the same connection operation can be used for all these cases. The resulting model \refEq{eqn:ConnectSystems_Model} is obtained by applying \refEq{eqn:ConnectSystems_ABCD}.
\item
\emph{Creating control loops.} Connecting a controller, that is, a building block type as given by \refEq{eqn:General_Controller}, to a converter is done by creating an open-loop model \refEq{eqn:ConverterController_General} using \refEq{eqn:ConverterController_OpenLoop}. For closing the control loop, the model's $\Matrix{A}$ matrix must be adapted using \refEq{eqn:ConverterController_ClosedLoop_A}, with a feedback gain matrix $\Matrix{K}$ chosen depending on the controlled variable, for example, one of  \refEq{eqn:K_CurrentMode} or \refEq{eqn:K_VoltageMode} for current or voltage mode control.
\end{enumerate}

All resulting models (connected subsystems, open- or closed-loop models) are of the same structure as the generic model \refEq{eqn:TwoPort_Controlled}, which means that frequency-domain characteristics as listed in the following section can be extracted at any modeling stage.

Since the structure and linearity of the resulting state-space models are preserved by the connection operations, there is no impact on the small-signal modeling accuracy of the building blocks.

% -----------------------------------------------------------------------------

\subsection{Frequency-Domain Analysis}
\label{sec:Summary_TF}

As mentioned before, one major benefit of the modular state-space modeling approach presented in this article is that, when needed, frequency-domain transfer functions can easily be obtained at any modeling stage.
\refTableBegin{tab:TransferFunctions} lists all frequency-domain models that can be extracted from the generic converter or converter system model \refEq{eqn:TwoPort_Controlled}, regardless of the current modeling state (e.g., with or without closed-loop control, with or without filters and loads etc.).

\begin{table}[h]
\centering
\begin{threeparttable}
\caption{Transfer functions than can be obtained from the converter (system) model \refEq{eqn:TwoPort_Controlled}, cf.\ Section \ref{sec:General_Converter}, at any modeling stage (e.g., with or without controller).}
\label{tab:TransferFunctions}
\begin{tabular}{llll}
\toprule
\textbf{Name}  &  \multicolumn{3}{l}{\textbf{Definition}}  \\
\midrule
Control-to-output \tnote{$\dagger$}
&  $G_\mathrm{co}(s)$
&  $= \D\frac{v_\mathrm{out}(s)}{\mathit{ctl}(s)}$
&  $= \D\Matrix{C}_2 \cdot \left( s \Matrix{I} - \Matrix{A} \right)^{-1} \cdot \Matrix{B}_3 + D_{23}$
\\
Output impedance
&  $Z_\mathrm{out}(s)$
&  $= \D\frac{v_\mathrm{out}(s)}{i_\mathrm{out}(s)}$
&  $= \D\Matrix{C}_2 \cdot \left( s \Matrix{I} - \Matrix{A} \right)^{-1} \cdot \Matrix{B}_2 + D_{22}$
\\
Input admittance
&  $Y_\mathrm{in}(s)$
&  $= \D\frac{i_\mathrm{in}(s)}{v_\mathrm{in}(s)}$
&  $= \D\Matrix{C}_1 \cdot \left( s \Matrix{I} - \Matrix{A} \right)^{-1} \cdot \Matrix{B}_1 + D_{11}$
\\
Forward voltage gain
&  $G_\mathrm{v}(s)$
&  $= \D\frac{v_\mathrm{out}(s)}{v_\mathrm{in}(s)}$
&  $= \D\Matrix{C}_2 \cdot \left( s \Matrix{I} - \Matrix{A} \right)^{-1} \cdot \Matrix{B}_1 + D_{21}$
\\
Reverse current gain
&  $G_\mathrm{i}(s)$
&  $= \D\frac{i_\mathrm{in}(s)}{i_\mathrm{out}(s)}$
&  $= \D\Matrix{C}_1 \cdot \left( s \Matrix{I} - \Matrix{A} \right)^{-1} \cdot \Matrix{B}_2 + D_{12}$
\\
\bottomrule\addlinespace[\belowrulesep]
\end{tabular}
\begin{tablenotes}
\footnotesize
\item[$\dagger$] If there is more than one control input (after connecting several converters to one model), the control-to-output transfer function for the $k$-th control input is obtained using the $\Matrix{B}_{2+k}$ column of the $\Matrix{B}$ matrix and $D_{2,(2+k)}$ instead of $\Matrix{B}_3$ and $D_{23}$.
\end{tablenotes}
\end{threeparttable}
\end{table}

% -----------------------------------------------------------------------------
% Examples
% -----------------------------------------------------------------------------

\section{Examples}
\label{sec:Examples}

In the following, three modular modeling examples are presented with special emphasis on demonstrating the systematic methods to extend two-port oriented state-space models of converters with control loops (method from Section \ref{sec:ConverterController}, example in Section \ref{sec:Examples_Multiloop}) and to create an overall model from series-connections of subsystems (method from Section \ref{sec:Connecting}, examples in Sections \ref{sec:Examples_BoostFilter} and \ref{sec:Examples_BoostBuck}).

It should be noted that the parameterization of converters and controllers in this section was chosen rather arbitrarily and solely to serve as numerical examples for the modeling methods developed here.
For simplicity, all inductors and capacitors in the following examples are being modeled with a uniform equivalent series resistance (ESR) value of $r_\mathrm{L/C} = \SI{10}{\milli\ohm}$, unless otherwise noted.

As will be demonstrated, all models developed in these examples can be constructed from already available modules, thus no manual effort is required regarding the derivation of state-space equations.
In order to illustrate the validity of the models in an intuitive manner, comparisons will be provided with time-domain results from circuit simulations.

% -----------------------------------------------------------------------------

\subsection{Buck Converter with Multiloop (I2) Control}
\label{sec:Examples_Multiloop}

The systematic approach to extend a converter model by attaching a control loop (presented in Section \ref{sec:ConverterController}) opens up the possibility of studying characteristics of converters with multiple control loops in an easy manner.
Therefore, as a first example, a buck converter in peak current mode control will be examined, which will be extended by a second control loop for the average current (also known as I2 control \cite{Yan:2014}) and a third outer voltage control loop.
Only three steps are necessary to create the complete model from already available building blocks:
\begin{enumerate}
\item
As the starting point to build up the complete model, the two-port state-space model for the PCM buck converter from Reference \cite{Smithson:2015} is being used.
The example is parameterized as given in the following.
Operating point:
$V_\mathrm{in} = \SI{24}{\volt}$, $V_\mathrm{out} = \SI{12}{\volt}$, $I_\mathrm{out} = \SI{2.4}{\ampere}$.
Power stage:
$f_\mathrm{Switch} = \SI{50}{\kilo\hertz}$, $L = \SI{100}{\micro\henry}$, $C = \SI{100}{\micro\farad}$.
Peak current mode controller: external ramp of $\SI{1.5}{\ampere}$ per switching~cycle.

\item
For the I2 current loop, an additional controller for the average inductor current is being added.
A Type~1 controller with $K_\mathrm{i} = 20$,$000$ is employed, cf.\ Appendix \ref{sec:Controller}, which is connected to the state-space model as described in Section \ref{sec:ConverterController_CurrentMode}.

\item
For the outer voltage control loop, a Type~2 controller with zero at $\SI{0.3}{\kilo\hertz}$, pole at $\SI{25}{\kilo\hertz}$ and $K_\mathrm{i} = 3000$ is being used, cf.\ Appendix \ref{sec:Controller}.
It is connected to the model as described in Section \ref{sec:ConverterController_VoltageMode}.
\end{enumerate}

\begin{figure}[h]
    \centering%
    \includegraphics{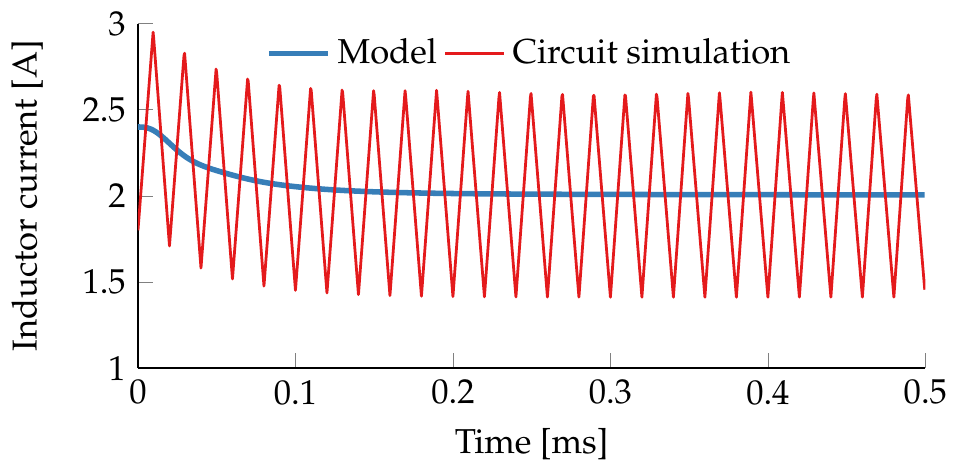}%
    \caption{Impact of a reference step $\SI{2.4}{\ampere} \rightarrow \SI{2.0}{\ampere}$ at $t = \SI{0}{\second}$ on the inductor current of the buck converter from example Section \ref{sec:Examples_Multiloop} with closed I2-current loop.}
    \label{fig:example_Buck_I2_StepRef}
\end{figure}

To validate the model in the time domain, it is being compared against a circuit simulation of this converter/controller configuration with three test cases.
Firstly, a reference step of the I2 current control loop (i.e., using the model after step 2 as given above) is simulated using both the detailed circuit simulation and the transfer function for the control loop of the average inductor current, which can easily be obtained from the state-space model (transfer function from the control input to the state variable of the inductor current).
The results are given in \refFig{fig:example_Buck_I2_StepRef} and confirm that the dynamics of the average inductor current under a combined average and peak current mode (I2) control are very well captured by the state-space model.

\begin{figure}[h]
    \centering%
    \includegraphics{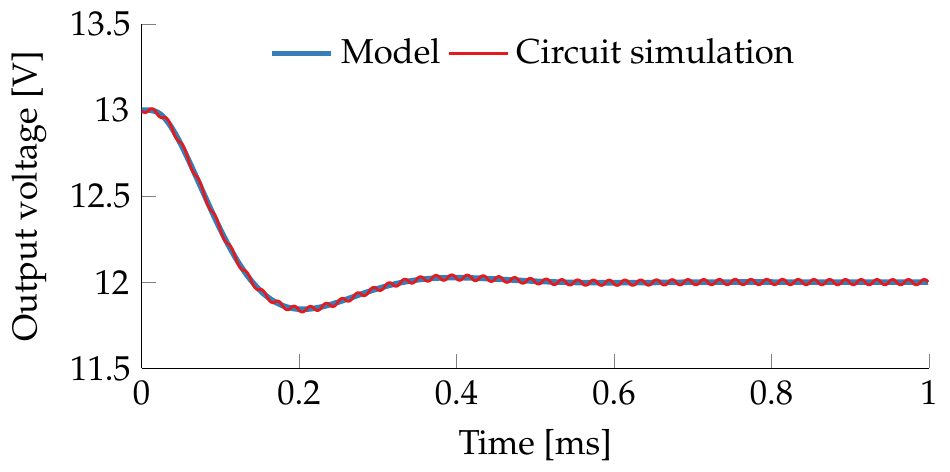}%
    \caption{Reference step $\SI{13}{\volt} \rightarrow \SI{12}{\volt}$ at $t = \SI{0}{\second}$ of the buck converter from example Section \ref{sec:Examples_Multiloop} with closed voltage loop and underlying I2 current control.}
    \label{fig:example_Buck_I2_V_StepRef}
\end{figure}

As a second test, a reference voltage step is performed using the outer voltage control loop (i.e., with the complete model after step 3).
The results are given in \refFig{fig:example_Buck_I2_V_StepRef} and, again, the circuit simulation and state-space model agree very well.

\begin{figure}[h!]
    \centering%
    \includegraphics{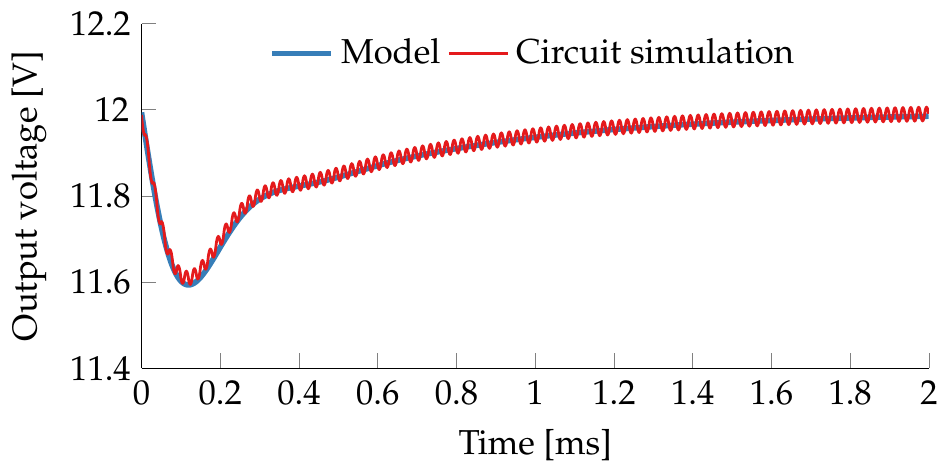}%
    \caption{Impact of a 25\%\ load step $\SI{2.4}{\ampere} \rightarrow \SI{3.0}{\ampere}$ at $t = \SI{0}{\second}$ on the output voltage of the buck converter from example Section \ref{sec:Examples_Multiloop} with closed voltage loop.}
    \label{fig:example_Buck_I2_V_StepLoad}
\end{figure}

Finally, +25\% load step is being applied.
The results in \refFig{fig:example_Buck_I2_V_StepLoad} once more confirm that the state-space model captures the dynamics of the converter under this multiloop control configuration very well.

Again, it should be noted that frequency-domain characteristics such as impedance plots or control-to-output transfer functions are all readily available at any stage of the state-space modeling procedure, cf.\ \refTable{tab:TransferFunctions} but omitted here for brevity.
There is no necessity anymore to derive custom models or transfer functions for the I2 current loop configuration, in contrast to existing work \cite{Yan:2014,He:2016,Li:2016}.

% -----------------------------------------------------------------------------

\subsection{Boost Converter with Input Filter}
\label{sec:Examples_BoostFilter}

With the model connection approach presented in Section \ref{sec:Connecting}, building up higher-order state-space models has become an easy task.
As an example, the impact of additional input circuitry on a voltage mode controlled boost converter will be studied here.
Four steps are necessary in order to obtain the final model from already available modules:
\begin{enumerate}
\item
A linearized state-space model for a boost converter in CCM as given in Appendix \ref{sec:Converter_Example_Boost}.
The~example is parameterized as given in the following.
Operating point:
$V_\mathrm{in} = \SI{10}{\volt}$, $V_\mathrm{out} = \SI{24}{\volt}$, $I_\mathrm{out} = \SI{1.2}{\ampere}$.
Power stage:
$f_\mathrm{Switch} = \SI{100}{\kilo\hertz}$, $L = \SI{20}{\micro\henry}$, $C = \SI{220}{\micro\farad}$.

\item
A load resistance (model from Appendix \ref{sec:Passive_R}) corresponding to the operating point is connected at the converter output using Section \ref{sec:Connecting}.

\item
A Type~3 controller (cf.\ Appendix \ref{sec:Controller}) is being employed for the voltage control loop, with zeros placed at $\SI{10}{\kilo\hertz}$, poles at $\SI{0.1}{\kilo\hertz}$ and $\SI{50}{\kilo\hertz}$, and $K_\mathrm{i} = 10$.
It is connected to the model as described in Section \ref{sec:ConverterController_VoltageMode}.

\item
Finally, an LC filter model (cf.\ Appendix \ref{sec:Passive_LC}) accounting for both an input filter and wiring is connected at the input of the converter model using Section \ref{sec:Connecting}, with $L = \SI{5}{\micro\henry}$, $r_\mathrm{L} = \SI{50}{\milli\ohm}$, and $C = \SI{1}{\micro\farad}$.
\end{enumerate}

\begin{figure}[h]
    \centering%
    \includegraphics{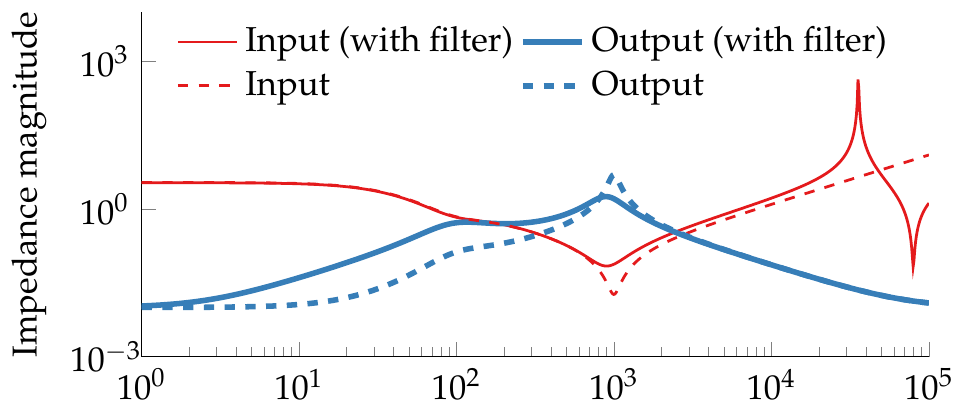}%
    \\
    \includegraphics{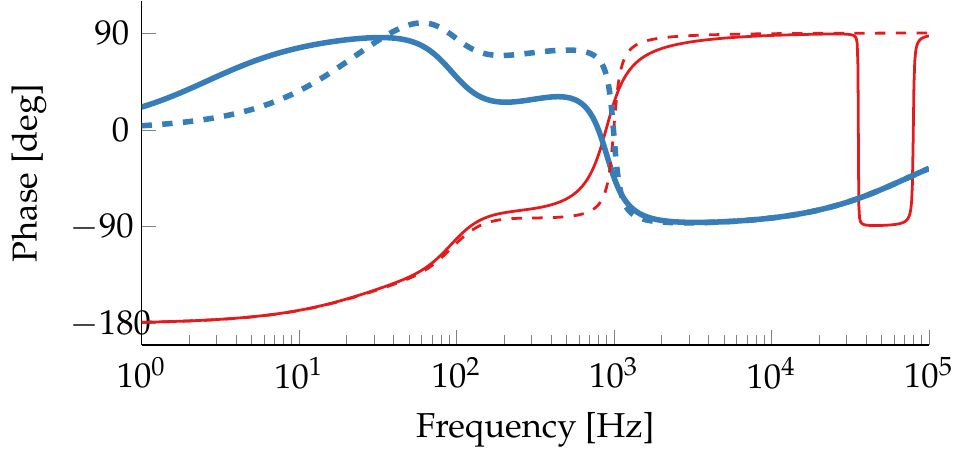}%
    \caption{Input and output impedance of the boost converter from example Section \ref{sec:Examples_BoostFilter}, with closed voltage loop, both with input filter (solid lines) and without input filter (dashed lines).}
    \label{fig:example_Boost_Impedance}
\end{figure}

As a first step, the influence of the input filter on the frequency-domain characteristics will be examined.
The input and output impedance transfer functions can easily be obtained from our two-port oriented state-space model, cf.\ \refTable{tab:TransferFunctions}.
\refFigBegin{fig:example_Boost_Impedance} shows the impedance magnitude and phase plots for the voltage mode controlled boost converter both with and without input filter.
In the input impedance plots one can clearly notice the negative resistance behavior of the controlled converter at low frequencies.
Furthermore one can, for instance, observe a dampening effect of the input filter around $\SI{1}{\kilo\hertz}$ in the output impedance, which has an impact on load step dynamics.

\begin{figure}[h]
    \centering%
    \includegraphics{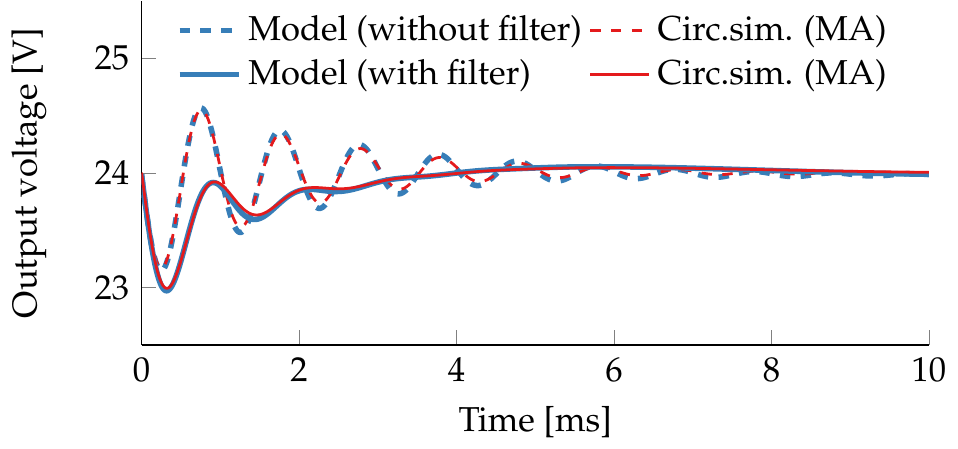}%
    \caption{Impact of a 100\% load step $\SI{1.2}{\ampere} \rightarrow \SI{2.4}{\ampere}$ at $t = \SI{0}{\second}$ on the output voltage of the boost converter from example Section \ref{sec:Examples_BoostFilter} with closed voltage loop, both with input filter (solid lines) and without input filter (dashed lines). For better comparability, the circuit simulation results were filtered (moving average) to remove the switching ripple voltage.}
    \label{fig:example_Boost_StepLoad}
\end{figure}

Therefore, a time-domain simulation of a +100\% load step at the output will be performed with and without input filter and compared against the results of detailed time-domain circuit simulations.
The results in \refFig{fig:example_Boost_StepLoad} not only confirm the good accuracy of the state-space model, but also clearly demonstrate the dampening effect of this particular input circuitry on the load step dynamics, as predicted in the frequency-domain analysis.

% -----------------------------------------------------------------------------

\subsection{Series-Connected Boost and Buck Converter Stages}
\label{sec:Examples_BoostBuck}

\begin{figure}[h]
    \centering%
    \includegraphics{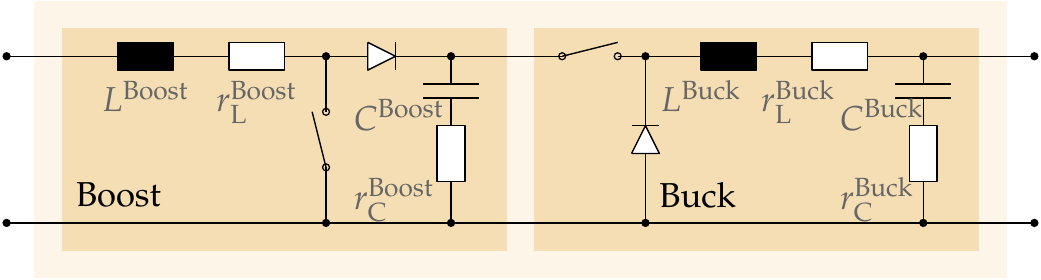}%
    \caption{Two-port view of a series connection of boost and buck converter stages. In both stages, the filter elements $L$ and $C$ are being modeled with equivalent series resistances (ESR).}
    \label{fig:TwoPort_BoostBuck}
\end{figure}

Since the model connection approach developed in Section \ref{sec:Connecting} is not restricted to attaching subsystems with passive components, one can just as easily connect models of controlled converters in a source/load configuration.
Therefore, as a final example, we will create a series-connected configuration of the boost and buck stages examined in Sections \ref{sec:Examples_Multiloop} and \ref{sec:Examples_BoostFilter}, respectively, cf.\ \refFig{fig:TwoPort_BoostBuck}.
The following (almost trivial) steps are required to obtain the complete model:
\begin{enumerate}
\item
Reuse the final model from Section \ref{sec:Examples_BoostFilter} (voltage-controlled boost converter including input filter) but without the load resistance, since the buck converter will constitute the new load to the boost~stage.

\item
Reuse the final model from Section \ref{sec:Examples_Multiloop} (voltage-controlled buck converter with underlying I2 current control).

\item
Connect the buck converter model at the output of the boost converter, as described in Section \ref{sec:Connecting}.
\end{enumerate}

As a small example of the possibilities offered by the all-encompassing state-space model, the interactions between the two power stages will be examined in the following test case.
Again, a +25\% load step is applied at the output of the buck stage and the impact on the intermediate output voltage of the boost stage is observed.
From the state-space model, a transfer function from the output current (i.e., the load) to the state variable representing the output capacitor voltage of the boost stage can easily be extracted.
It is, once more, being compared against a time-domain circuit simulation containing the series-connected converter configuration.

\begin{figure}[h]
    \centering%
    \includegraphics{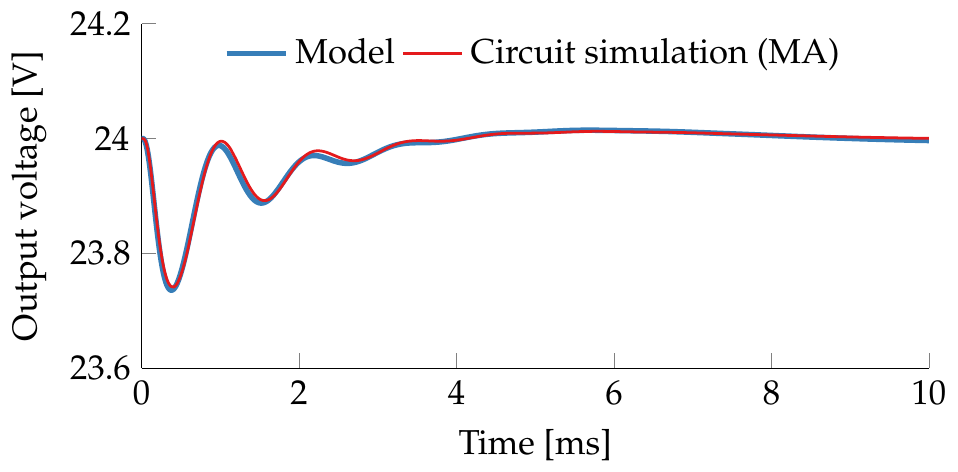}%
    \caption{Impact of a 25\% output load step $\SI{2.4}{\ampere} \rightarrow \SI{3.0}{\ampere}$ at $t = \SI{0}{\second}$ of the series-connected buck stage from example Section \ref{sec:Examples_BoostBuck} on the intermediate output voltage of the boost stage. For better comparability the circuit simulation result was filtered (moving average) to remove the switching ripple voltage.}
    \label{fig:example_BoostBuck_StepLoad}
\end{figure}

From the results in \refFig{fig:example_BoostBuck_StepLoad} one can conclude that the overall state-space model is very well able to predict the essential dynamics.
Since the buck stage has much faster dynamics than the boost stage, the load step dynamics in \refFig{fig:example_BoostBuck_StepLoad} closely resemble \refFig{fig:example_Boost_StepLoad} in this case but it has to be emphasized again that the results in \refFig{fig:example_BoostBuck_StepLoad} fully incorporate small-signal interactions of the two connected power stages with all their respective controllers.

% -----------------------------------------------------------------------------
% Conclusion
% -----------------------------------------------------------------------------

\section{Conclusions}
\label{sec:Conclusion}

The rederivation of state-space models of converter systems when adding input filters or non-trivial loads such as other converters was, up to now, considered a painful process \cite{Basso:2014}---and rightfully so.
With the systematic approach of extending such state-space models presented in this article, adding control loops and building up system models from series-connected modules such as filters, loads, or converters have become straightforward tasks.

This modular approach leverages already existing models for common converter or filter topologies by incorporating them as building blocks, such that users do not have to manually derive larger sets of state-space equations anymore in many application cases.
If not yet available, the~modeling efforts required are only limited to the respective subsystem.

The small-signal modeling accuracy of the building blocks is not compromised by the proposed connection operations, the resulting models in all modeling stages preserve their structure and always remain linear.

The validity of the resulting small-signal models has been illustrated and compared in the time-domain against circuit simulations.

At all stages of the modeling procedure proposed in this article, the benefits of a two-port oriented state-space description become obvious, since input/output impedance and control-to-output or control-to-state transfer functions are easily obtainable from one single model.
This also allows for the design of controllers starting from the innermost loop in a systematic fashion, while incorporating interactions with non-trivial source or load elements.

% -----------------------------------------------------------------------------
% -----------------------------------------------------------------------------
% Appendix
% -----------------------------------------------------------------------------
% -----------------------------------------------------------------------------

\appendix

% -----------------------------------------------------------------------------
% Passive Components
% -----------------------------------------------------------------------------

\section{Building Blocks: Passive Component Models}
\label{sec:Passive}

The generic two-port modeling approach for passive subsystems of a converter system has already been introduced in Section \ref{sec:General_Passive}.
In this section, two typical examples often found in a converter system will be given: an (ideal) resistive load, and an LC input filter.
These are also used in the examples presented in Section \ref{sec:Examples}.

% -----------------------------------------------------------------------------

\subsection{Resistive Load}
\label{sec:Passive_R}
With no storage elements, there are no dynamics in a resistive system as shown in Figure \ref{fig:TwoPort_Passive}a, which means that $\Matrix{A} = 0$, $\Matrix{B} = 0$, $\Matrix{C} = 0$ hold. Only the $\Matrix{D}$ matrix elements of a two-port state-space model according to \refEq{eqn:TwoPort_Passive} are present, and one obtains the model \refEq{eqn:TwoPort_Passive_R}.

\begin{equation}
\label{eqn:TwoPort_Passive_R}
\begin{pmatrix}
i_\mathrm{in}(t)
\\
v_\mathrm{out}(t)
\end{pmatrix}
=
\begin{pmatrix}
\frac{1}{R}  &  -1
\\
1  &  0
\end{pmatrix}
\cdot
\begin{pmatrix}
v_\mathrm{in}(t)
\\
i_\mathrm{out}(t)
\end{pmatrix}
\end{equation}

% -----------------------------------------------------------------------------

\subsection{LC Filter}
\label{sec:Passive_LC}
\begin{figure}[h]
    \centering%
    \subfloat[]{%
        \includegraphics{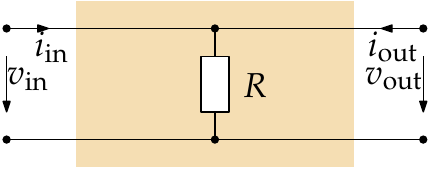}%
        \label{fig:TwoPort_Passive_R}%
    }
    \hspace{2cm}%
    \subfloat[]{%
        \includegraphics{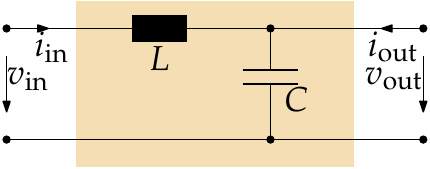}%
        \label{fig:TwoPort_Passive_LC}%
    }
    \caption{Two-port network subsystems for passive elements. {(\textbf{a})} Resistive load. {(\textbf{b})} LC filter.}
    \label{fig:TwoPort_Passive}
\end{figure}

In \refEq{eqn:TwoPort_Passive_LC}, the two-port state-space model of an LC filter as depicted in Figure \ref{fig:TwoPort_Passive}b is being given.
In~addition to the circuit in Figure \ref{fig:TwoPort_Passive}b, the effects of equivalent series resistances (ESR) in both inductor (ESR $r_\mathrm{L}$) and capacitor (ESR $r_\mathrm{C}$) have been accounted for in \refEq{eqn:TwoPort_Passive_LC}.

\begin{equation}
\label{eqn:TwoPort_Passive_LC}
\begin{split}
\begin{pmatrix}
\dot{i}_\mathrm{L}(t)  \\  \dot{v}_\mathrm{C}(t)
\end{pmatrix}
=&
\begin{pmatrix}
-\frac{r_\mathrm{L} + r_\mathrm{C}}{L}  &  -\frac{1}{L}
\\[1ex]
\frac{1}{C}  &  0
\end{pmatrix}
\cdot
\begin{pmatrix}
i_\mathrm{L}(t)  \\[1ex]  v_\mathrm{C}(t)
\end{pmatrix}
+
\begin{pmatrix}
\frac{1}{L}  &  -\frac{r_\mathrm{C}}{L}
\\[1ex]
0  &  \frac{1}{C}
\end{pmatrix}
\cdot
\begin{pmatrix}
v_\mathrm{in}(t)  \\[1ex]  i_\mathrm{out}(t)
\end{pmatrix}
\\
\begin{pmatrix}
i_\mathrm{in}(t)  \\  v_\mathrm{out}(t)
\end{pmatrix}
=&
\begin{pmatrix}
1  &  0
\\
r_\mathrm{C}  &  1
\end{pmatrix}
\cdot
\begin{pmatrix}
i_\mathrm{L}(t)  \\  v_\mathrm{C}(t)
\end{pmatrix}
+
\begin{pmatrix}
0  &  0
\\
0  &  r_\mathrm{C}
\end{pmatrix}
\cdot
\begin{pmatrix}
v_\mathrm{in}(t)  \\  i_\mathrm{out}(t)
\end{pmatrix}
\raisetag{11ex}
\end{split}
\end{equation}

As a side note it can be stated that manually creating a model for a two-stage LC filter is not necessary.
A benefit of the connection method presented in Section \ref{sec:Connecting} will be that one can create a series connection of two LC filter models from \refEq{eqn:TwoPort_Passive_LC} if that is necessary.

More generally speaking: it is beneficial to create very simple building blocks as demonstrated in this section, and bypass the error-prone process of manually modeling a larger system by creating a complete system model from the combination of these small subsystems instead.
The necessary tools are presented in Section \ref{sec:Connecting}.

% -----------------------------------------------------------------------------
% Converter Modeling
% -----------------------------------------------------------------------------

\section{Building Blocks: Converter Models}
\label{sec:Converter}

% -----------------------------------------------------------------------------
\vspace{-6pt}

\subsection{Available Converter Models}
\label{sec:Converter_Literature}

The two-port state-space approach to converter modeling goes beyond most state-space models found in the literature, which mostly do not account for the output current $i_\mathrm{out}(t)$ as an additional input variable of the model.
This is probably due to the fact that state-space models have so far mostly been used only as an intermediate step in order to derive transfer functions in frequency domain.

Nevertheless, a variety of two-port oriented state-space models can already be found in more recent work, and these can be used here almost in a plug-and-play fashion.
For example, models for buck, boost, and flyback converter in peak current mode (PCM) control, all in continuous conduction mode (CCM), are presented in \cite{Smithson:2015}; and PCM-controlled buck and boost converters in DCM in \cite{Suntio:2018:Energies,Suntio:2019:Energies}, respectively.
Furthermore, ready-to-use state-space models for buck, boost, and buck-boost converters can be found in \cite{Suntio:2009,Suntio:2017}, which have to be seen as the most comprehensive references devoted to systematic two-port modeling of converters.

In this article, the recommended strategy is to model the load subsystem of the converter separately (cf.\ Appendix \ref{sec:Passive}), and connect it to the converter models afterwards (cf.\ Section \ref{sec:Connecting}).
This implies that, ideally, the converter modeled with \refEq{eqn:TwoPort_Controlled} should be unterminated at first.
However, most state-space converter models found in the literature do feature a resistive load, but these can be used here by letting $R \rightarrow \infty$, as done e.g., in \cite{Suntio:2006}.
The models presented in \cite{Suntio:2009} do already describe unterminated converters.

% -----------------------------------------------------------------------------

\subsection{Example: Boost Converter}
\label{sec:Converter_Example_Boost}

\begin{figure}[h]
    \centering%
    \includegraphics{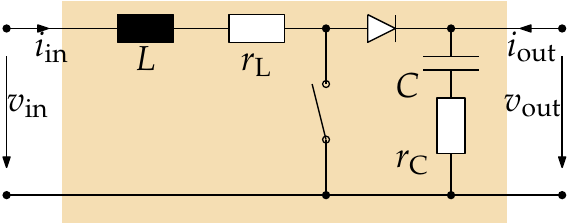}%
    \caption{Two-port view of an unterminated boost converter stage.}
    \label{fig:TwoPort_Boost}
\end{figure}

For an unterminated boost converter operating in CCM (cf.\ \refFig{fig:TwoPort_Boost}), the $\Matrix{A}$, $\Matrix{B}$, $\Matrix{C}$, $\Matrix{D}$ matrices for a linearized state-space model \refEq{eqn:TwoPort_Controlled} with small simplifications (e.g., ideal switches) are given in \refEq{eqn:TwoPort_Converter_Boost_ABCD}.
The power stage inductor $L$ and capacitor $C$ are being modeled with ESR values $r_\mathrm{C}$ and $r_\mathrm{L}$.
$D$, $I_\mathrm{L}$, and $V_\mathrm{out}$ define the operating point for the duty cycle, the average inductor current, and the output voltage, respectively.
The two state variables are the inductor current $i_\mathrm{L}(t)$ and the output capacitor voltage $v_\mathrm{C}(t)$, i.e., $\Vector{x}(t) = \begin{pmatrix} i_\mathrm{L}(t) & v_\mathrm{C}(t) \end{pmatrix}^\Transpose$.

\begin{equation}
\label{eqn:TwoPort_Converter_Boost_ABCD}
\begin{aligned}
&\Matrix{A}
=
\begin{pmatrix}
-\D\frac{((1-D) \cdot r_\mathrm{C} + r_\mathrm{L})}{L}  &  -\D\frac{1-D}{L}
\\[2ex]
\D\frac{1-D}{C}  &  0
\end{pmatrix}
,
& \quad &
\Matrix{B}
=
\begin{pmatrix}
\D\frac{1}{L}  &  -\D\frac{(1-D) \cdot r_\mathrm{C}}{L}  &  \D\frac{V_\mathrm{out}}{L}
\\[2ex]
0  &  \D\frac{1}{C}  &  -\D\frac{I_\mathrm{L}}{C}
\end{pmatrix}
,
\\[1em]
&\Matrix{C}
=
\begin{pmatrix}
1  &  0
\\
(1-D) \cdot r_\mathrm{C}  &  1
\end{pmatrix}
,
& \quad &
\Matrix{D}
=
\begin{pmatrix}
0  &  0  &  0
\\
0  &  r_\mathrm{C}  &  -r_\mathrm{C} \cdot I_\mathrm{L}
\end{pmatrix}
\end{aligned}
\end{equation}

This model is also used and compared against time-domain circuit simulations in the examples presented in Section \ref{sec:Examples_BoostFilter} (boost converter with input filter) and Section \ref{sec:Examples_BoostBuck} (series-connected boost and buck stages).

% -----------------------------------------------------------------------------
% Controller Modeling
% -----------------------------------------------------------------------------

\section{Building Blocks: Controller Models}
\label{sec:Controller}

For the control of voltage and current in power converters, some of the most prominent examples of controllers used in practice besides proportional-integral controllers are the so-called Type~1, Type~2, and Type~3 controllers described in \cite{Venable:TP3,Basso:2012}.
These are integral controllers with zero, one, or two lead-lag elements, respectively, and are also being used in the examples in Section \ref{sec:Examples}.

In \refTable{tab:ControllerType123}, transfer functions and state-space models according to \refEq{eqn:General_Controller} are given for these controller types.
For all of them, the state-space models do not possess direct feedthrough terms in \refEq{eqn:General_Controller}, i.e., $\Matrix{D}_\mathrm{C} = 0$.

\begin{table}[h]
\caption{Transfer function $G_\mathrm{C}(s)$ and state-space model matrices $\Matrix{A}_\mathrm{C}$, $\Matrix{B}_\mathrm{C}$, $\Matrix{C}_\mathrm{C}$ according to \refEq{eqn:General_Controller} for controllers known as Type 1, 2, and 3.}
\label{tab:ControllerType123}
\centering
\begin{tabular}{lll}
\toprule
\textbf{Type 1}  &  \textbf{Type 2 } &  \textbf{Type 3}\\
\midrule
$G_\mathrm{C}(s)
= \D\frac{K_\mathrm{i}}{s}
$
&
$G_\mathrm{C}(s)
= \D\frac{K_\mathrm{i}}{s} \cdot \frac{ 1 + T_\mathrm{z} s }{ 1 + T_\mathrm{p} s }$
&
$G_\mathrm{C}(s)
= \D\frac{K_\mathrm{i}}{s} \cdot \frac{ ( 1 + T_\mathrm{z1} s ) ( 1 + T_\mathrm{z1} s ) }{  ( 1 + T_\mathrm{p1} s )  ( 1 + T_\mathrm{p2} s ) }$
\\\midrule
$\Matrix{A}_\mathrm{C} = 0$
&
$\Matrix{A}_\mathrm{C} = \begin{pmatrix}
0  &  0
\\[1ex]
1  &  -\frac{1}{T_\mathrm{p}}
\end{pmatrix}$
&
$\Matrix{A}_\mathrm{C} = \begin{pmatrix}
0  &  0  &  0
\\[1ex]
1  &  0  &  -\frac{1}{T_\mathrm{p1} T_\mathrm{p2}}
\\[1ex]
0  &  1  &  -\frac{T_\mathrm{z1} + T_\mathrm{z2}}{T_\mathrm{p1} T_\mathrm{p2}}
\end{pmatrix}$
\\\midrule
$\Matrix{B}_\mathrm{C} = K_\mathrm{i}$
&
$\Matrix{B}_\mathrm{C} = \begin{pmatrix}
\frac{K_\mathrm{i}}{T_\mathrm{p}}
\\[1ex]
\frac{K_\mathrm{i} T_\mathrm{z}}{T_\mathrm{p}}
\end{pmatrix}$
&
$\Matrix{B}_\mathrm{C} = \begin{pmatrix}
\frac{K_\mathrm{i}}{T_\mathrm{p1} T_\mathrm{p2}}
\\[1ex]
\frac{K_\mathrm{i} \left( T_\mathrm{z1} + T_\mathrm{z2} \right) }{T_\mathrm{p1} T_\mathrm{p2}}
\\[1ex]
\frac{K_\mathrm{i} T_\mathrm{z1} T_\mathrm{z2}}{T_\mathrm{p1} T_\mathrm{p2}}
\end{pmatrix}$
\\\midrule
$\Matrix{C}_\mathrm{C} = 1$
&
$\Matrix{C}_\mathrm{C} = \begin{pmatrix}
0  &  1
\end{pmatrix}$
&
$\Matrix{C}_\mathrm{C} = \begin{pmatrix}
0  &  0  &  1
\end{pmatrix}$
\\
\bottomrule
\end{tabular}
\end{table}

% -----------------------------------------------------------------------------
% -----------------------------------------------------------------------------

\end{document}